\documentstyle[11pt]{article}
\begin{document}
\vspace{3mm}
\begin{center}
{\Large \bf The non-local action for the induced 2d supergravity.}
\footnote { Work supported in part by Armenian Foundation of Scientific
Researches and by grant 211-5291 YPI of the German Bundesministerium
f\"ur Forschung und Technologie, Federal Republic of Germany}\\
\vspace{1cm}
{\large D.R.Karakhanyan}\\ 
{\em Yerevan Physics Institute, Republic of Armenia \\
(Alikhanian Brothers St. 2, Yerevan 375036, Armenia)\\
E-mail: karakhan@lx2.yerphi.am}\\
\end{center}
\begin{abstract} 
The two-dimensional simple supergravity is reexamined from
the point of view of super-Weyl group cohomologies. The
non-local form of the effective action of 2d supergravity 
which generalise the famous $R\frac{1}{\Box}R$ is obtained.
\end{abstract}
\section{ Introduction}

The two-dimensional field-theoretical models are widely investigated at
last two decades. The $2d$-models provides us an excellent laboratory for
studying realistic, but much more complicated four-dimensional models.
Indeed such phenomena as renormalization, asymptotic freedom, dimensional 
transmutation and others hold in integrable $2d$ models and can be described 
exactly. Finally from $2d$ models we can guess some information about
structure of corresponding $4d$-models.

The $2d$-gravity takes an apparent place among $2d$-models. It happens not
only by the fact that gravity is most mysterious part of quantum field 
theory, but also by its close relation to the string theory.
After the famous work of A.M.Polyakov \cite{p1}, which reduces non-critical 
strings to induced gravity, that theory became intensively studied,
A.M.Polyakov \cite{p2} was compute most important indices of the theory in 
light-cone formulation, F.Distler, H.Kawai and F.David \cite{ddk} was 
rederived that result in conformal gauge, then A.M.Polyakov, V.G.Kniznik and 
A.B.Zamolodchikov \cite{kpz} discover that anomaly cancellation implies that 
target space dimension has to be fractal. Later it was shown that different
regularization schemes are equivalent \cite{rrd}, it was found  formulation 
of $2d$ induced supergravity, computed in Weyl-invariant regularization
scheme too \cite{d}. The holomorphic properties of the effective action 
has been studied in \cite{ls} and in \cite{dk}. In \cite{ks} made attempt 
to understand extrinsic geometry relation to the induced gravity.
Authors of work \cite{fht} study relations between super-Virasoro and
super-Weyl anomaly and construct super-Weyl invariant functional by
adding non-local functional in order to cancell super-Virasoro anomaly.
But till now the non-local expression for $2d$ induced supergravity 
analogous to famous $R\frac{1}{\Box}R$ remains unknown. The present work 
is called to cover that missing.

\section { Computation}

The graviton modes in gravity are described by traceless excitations of
 metric. The invariance of an action under general 
coordinate transformations implies the independence of the action from 
such a variations of the metric field. As well known in $2d$ general 
covariance is accompanied by local Weyl-symmetry, which implies the 
independence of action from the remaining variations of the metric: 
$g^{\alpha\beta}\delta g_{\alpha\beta}$ too.
In other words, in $2d$ general covariant action actually is
independent from the metric and $2d$ space-time is locally flat.
However the quantum fluctuations spoil that symmetry and give rise the
conformal anomaly: the variation of the effective action under such  
variations of metric is no longer equal to zero
\begin {equation}
g^{\alpha\beta}\frac{\delta W}{\delta g_{\alpha\beta}}=R
\end {equation}
where $R$ is the curvature of $2d$ space-time.
It is easily to see that $R$ defined as variational derivative of $W$ 
satisfies to consistency condition: under Weyl rescalings 

\begin{equation}
\delta g_{\alpha\beta}=g_{\alpha\beta}\delta\sigma
\end{equation}
curvature transforms accordingly rule

\begin {equation}
\delta[\sqrt g R]=\sqrt g\Box\delta\sigma
\end {equation}
So the second derivative of $W$ by $\sigma$ is symmetric: 

\begin {equation}
\frac{\delta R(x)}{\delta\sigma(y)}=\frac{\delta R(y)}
{\delta\sigma(x)}.
\end {equation}
In $2d$ supergravity graviton's partner - gravitino describes 
by $spin\ 3/2$ components of Rarita-Schwinger field. So, local 
supersymmetry implies independents of the action under $spin\ 3/2$ 
variations of spin-vector field. Classically $2d$ supergravity also is 
locally flat: local supersymmetry in $2d$ is accompanied by invariance 
under variations of $spin\ 1/2$ part of gravitino too
\begin {equation}
\delta \chi_\alpha=\gamma_\alpha\delta\lambda
\end {equation}
That symmetry also is spoiled by quantum corrections
\begin {equation}
\frac{\delta W}{\delta\lambda}=\gamma_\alpha J^\alpha
\end {equation}
where $J^\alpha$ is supercurrent of theory (superpartner of 
stress-tensor). Now its contraction with $\gamma$ matrix is no longer 
equal to zero. In order to find super-Weyl anomaly, one can use integrability 
condition of the action: variational derivative of the curvature 
(Weyl-anomaly) by $\lambda$ should be equal to the super-Weyl anomaly's 
derivative by $\sigma$. The $\lambda$-dependence of curvature comes from 
non-minimal term of spin-connection:
\begin {equation}
\omega_\alpha=-e^a_\alpha\frac{\epsilon^{\mu\nu}}{e}\partial_\mu e^a_\nu-
2i\bar\chi_\alpha\gamma_5\gamma^\mu\chi_\mu
\end {equation}
Solving that equation, we can write down 
following set of anomaly equations for $2d$ supergravity computing in
super-coordinate invariant regularization:
\begin {eqnarray}
g^{\alpha\beta}T_{\alpha\beta}=R\\
\gamma^\alpha J_\alpha=-4i\frac{\epsilon^
{\alpha\beta}}{e}\gamma_5D_\alpha\chi_\beta
\end {eqnarray}
where $D_\alpha\lambda=\partial_\alpha\lambda+\frac{1}{2}\gamma_5\omega_
\alpha\lambda$. It is easely to check that last Wess-Zumino condition
is satisfied too, $\lambda$-derivative of super-Weyl anomaly (9) is
symmetric in sense of eq.(4). Hence eqs.(8) and (9) define the consistent 
multiplet of supertrace anomaly for 2d simple supergravity.  
This expression for anomaly coincide with result of authors
\cite{k-k}. Transforming r.h.s. of this equations by finite Weyl and
super-Weyl and multiplying them by $\delta\sigma$ and $\delta\lambda$
correspondingly and adding together we recognize the total variation of
the effective action, which is Neveu-Schwarz action plus $R\sigma$ and 
$-4i\frac{\epsilon^{\alpha\beta}}{e}\bar\lambda\gamma_5D_\alpha\chi_\beta$ 
terms.

\begin {eqnarray}
S(\sigma,\lambda;e^a_\alpha,\chi_\alpha)=\int ed^2x[R\sigma-4i\epsilon^{
\alpha\beta}/e\bar\lambda\gamma_5D_\alpha\chi_\beta-1/2g^{\alpha\beta}
\partial_\alpha\sigma\partial_\beta\sigma\\\nonumber
-i/2\bar\lambda\gamma^\alpha\partial_\alpha\lambda+i\bar\lambda\gamma^\beta
\gamma^\alpha\chi_\beta(\partial_\alpha\sigma-i/2\bar\lambda\chi_\alpha)]
\end {eqnarray}
This result has been obtained earlier in \cite{kk}. 
In agreement with general analyses of the Weyl anomaly and the 
reconstruction of the corresponding effective action \cite{mmk},
we suppose that the effective action has to be the sum of terms of the
form: $Anomaly($Weyl-inv.diff.op.$)^{-1}Anomaly$.
Now, the conformally invariant operators, which should be the
denominators in non-local expression for effective action, are 
defined as a variational derivatives of cocycle eq.(10) or what the 
same, of one of the Neveu-Schwarz action, because its differ by terms
linear with respect to $\lambda$ and $\sigma$. The cocycle eq.(10) is Weyl
co-boundary of the effective action $W[e^a_\alpha,\chi_\alpha]$, which we
are looking for. 

\begin {equation}
S(\sigma,\lambda;e^a_\alpha,\chi_\alpha)=W[e^{\sigma/2}e^a _\alpha,
e^{\sigma/4}(\chi_\alpha+\frac{1}{4}\gamma_\alpha\lambda)]-W[e^a_\alpha,\chi_\alpha]
\end {equation}
The cocyclic property of the $S$

\begin {equation}
S(\sigma_1+\sigma_2,\lambda_1+\lambda_2;e^a_\alpha,\chi_\alpha)=
S(\sigma_1,e^{-\sigma_2/4}\lambda_1;e^{\sigma_2/2}e^a_\alpha,
e^{\sigma_2/4}(\chi_\alpha+\frac{1}{4}\gamma_\alpha\lambda_2))
+S(\sigma_2,\lambda_2;e^a_\alpha,\chi_\alpha)
\end {equation}
now is trivial consequence of previous relation. However, multiplying  
those denominators by anomaly and jointing them together, one doesn't reach
desirable result, as easily to seen. Exploiting the fact that cocycle eq.(10)
takes quadratic form on super-Weyl group variables, one could tries to
perform gaussian integration in order to obtain the effective action.
Unfortunately this procedure leads to complicated expression, which seems
has in denominators differential operator degree greater than two. Apart
from that, it is very hard to prove relation eq.(11) for that expression.

In order to avoid those difficulties let us proceed by following.
First of all, let us notice that apart from curvature and the curl of 
Rarita-Schwinger field there are no terms with dimensions $2$ and $3/2$ whose 
are super Weyl invariant, satisfy to Wess-Zumino consistency condition and can 
be added to r.h.s. of eqs.(8) and (9) correspondingly . So as well as in 
ordinary 2d gravity there is no Weyl-invariant solutions of Wess-Zumino 
condition. Only term of dimension two can be considered
\begin {equation}
(\chi_\alpha\gamma^\beta\gamma^\alpha\chi_\beta)^2=(\frac{1}{2}\bar{\tilde\chi}
_\alpha\tilde\chi^\alpha)^2=0
\end {equation}
Indeed, that term can be considered as the fourth degree of $spin\ 3/2$ 
part of Rarita-Schwinger field $\tilde\chi_\alpha=\gamma^\beta\gamma_\alpha
\chi_\beta$ - quantity, which has only two independent components. 

Let us now "improve" our anomaly expressions. It is impossible to add
to such a local counterpart to $R$, in order to cancel it's $\sigma$-variation.
Rather, it is possible to combine $R$ in such a way to it's $\lambda$-variation  
vanish. Analogous improvement should to be done for super-Weyl anomaly.
So the expression for anomaly, convenient for us takes form:

\begin {eqnarray}
{\cal A}_2=R+i/2\nabla_\alpha(\bar\chi_\mu\gamma^\alpha\gamma^
\beta\gamma^\mu\chi_\beta)\\
{\cal 
A}_{3/2}=-4i\frac{\epsilon^{\alpha\beta}}{e}\gamma_5D_\alpha\chi_
\beta+D_\alpha(\gamma^\beta\gamma^\alpha\chi_\beta),\nonumber
\end {eqnarray}
Let us notice, that "improved" anomaly expression also satisfy to
Wess-Zumino condition, because improving terms can be represented
as the Weyl-coboundaries of the local functionals.

\begin {eqnarray}
\nabla_\alpha(\bar\chi_\mu\gamma^\alpha\gamma^\beta\gamma^\mu\chi_
\beta)(x)=-1/4\frac{\delta}{\delta\sigma(x)}\int ed^2y\bar\chi_\beta
\gamma^\beta D_\alpha\chi^\alpha \\
D_\alpha(\gamma^\beta\gamma^\alpha\chi_\beta)(x)=-2\frac{\delta}
{\delta\lambda(x)}\int ed^2y[\bar\chi_\mu\gamma^\nu\gamma^\mu
\gamma^\alpha D_\nu\chi_\alpha-i\bar\chi_\mu\gamma^\nu\gamma^\mu
\chi_\nu\bar\chi_\alpha\chi^\alpha]. \nonumber
\end {eqnarray}
Now, multiplying ${\cal A}_2$ by inverse Laplace operator and by
${\cal A}_2$ we notice, that it's coboundary is local expression, as
well as the coboundary of ${\cal A}_{3/2}({\cal D})^{-1}{\cal A}_{3/2}$,
here ${\cal D}=-i\gamma^\alpha D_\alpha+\chi_\alpha\bar\chi_\beta\gamma^
\alpha\gamma^\beta$ is kinetic operator of spinor fields in Neveu-
Schwarz action.
However, we obtain the extra "improving" terms in our coboundary.
Those are removed by adding the third non-local term to our action,
which is constructed from remaining conformally-invariant operator.
Finally the effective action takes the form
\begin {eqnarray}
W[e,\chi]=\int [e(R+i/2\nabla_\alpha(\bar\chi_\mu\gamma^\alpha\gamma
^\beta\gamma^\mu\chi_\beta))\frac{1}{e\Box}e(R+i/2\nabla_\alpha(\bar
\chi_\mu\gamma^\alpha\gamma^\beta\gamma^\mu\chi_\beta))+\nonumber\\
e^{3/4}(-4i\frac{\epsilon^{\alpha\beta}}{e}\gamma_5D_\alpha\chi_ 
\beta+D_\alpha(\gamma^\beta\gamma^\alpha\chi_\beta))\frac{1}{e^{1/2}
{\cal D}}e^{3/4}(-4i\frac{\epsilon^{\alpha\beta}}{e}\gamma_5D_\alpha
\chi_\beta+D_\alpha(\gamma^\beta\gamma^\alpha\chi_\beta))-\\
(R+i/2\nabla_\alpha(\bar\chi_\mu\gamma^\alpha\gamma^\beta\gamma^\mu
\chi_\beta))\frac{1}{e^{3/4}(i\gamma^\beta\gamma^\alpha\chi_\beta
\partial_\alpha)}e^{3/4}{\cal A}_{3/2}]\nonumber  
\end {eqnarray}
here $\Box$ is ordinary Laplace operator, acting on scalar fields.
This is our desirable expression for the action of induced $2d$
supergravity, computed in supercoordinat-invariant regularization.
The last term eq.(15) can be replaced by following local combination
\begin {equation}
\int ed^2x[\frac{i}{8}\bar\chi_\beta\gamma^\beta D_\alpha\chi^\alpha+
2(\bar\chi_\mu\gamma^\nu\gamma^\mu\gamma^\alpha D_\nu\chi_\alpha-i
\bar\chi_\mu\gamma^\nu\gamma^\mu\chi_\nu\bar\chi_\alpha\chi^\alpha)].
\end {equation} 
The anomaly equations (8) and (9) remain unchanged under that 
replacement, so the non-locality related to the last term of the
effective action (16) is removable. The addition of dimensionless 
actions - local countereterms like eq.(14) makes finite 
renormalization of the action. The overall constant in front of the 
action defined by the physical content of the theory.

The effective action for 2d supergravity, computed in super-Weyl
invariant regularization scheme can be obtained from (15) by adding
the non-super coordinate invariant action
\begin {equation}
S(\sigma,\lambda;e^a_\alpha,\chi_\alpha)|_{\sigma=\log e; \lambda=
-2\gamma^\beta\chi_\beta}
\end {equation}
accordingly to relation (11) it can be represented in manifest 
super-Weyl invariant form:
\begin {equation}
\tilde W[E^a_\alpha,\chi_\alpha]=W[e^{-1/2}e^a_\alpha,e^{-1/4}\gamma^
\beta\gamma^\alpha\chi_\beta]
\end {equation}
As easely to seen this action differs from eq.(16) by local expression'
hence can be reached from that by finite renormalization, which will
mean change of regularization scheme of the theory.

\section {Conclusion and outlook}

So, the main conclusion , that can be derived from our experience
of work with Weyl anomaly may be formulated as following statement.
The expressions acceptable as the conformal anomaly are following:
the density of the Euler characteristic, Weyl-invariant lagrangian
densities and variations of local functionals of appropriate 
dimensions (improving terms), those can be added to the action as the
local counterterms. There already exists such the representative of the
class of solutions of the consistency condition, which has the linear 
and diagonal (in supersymmetric case) finite Weyl-variation.
Corresponding cocycle will be quadratic. Integration of that cocycle
by group variables is easily to perform in order to obtain the 
non-local expression for the effective action.

The arguments described above can be also applied to the four-dimensional
supergravity. In that case, however, we cannot restore whole effective
action as in $2d$, because the anomaly equations contain information only
about anomalous dependence of the effective action from parameters of
super Weyl group - trace of the metric and $spin\ 1/2$ part of the  
Rarita-Schwinger field. After taking into account manifest general
covariance and local supersymmetry of the resulting expression there 
five components of the metric and eight components of Rarita-Schwinger 
field remain, dependence from which comes as an integration "constant" 
under integration of anomaly equations.

\section {Acknowledgments}
I would like to thank A.G.Sedrakyan, R.Kuriki and T. Hakobyan for 
useful notations. Also I thank the ICTP Trieste, where this paper
was finished, for kind hospitality.

~
\end {document}